# Temperature measurement in the Intel® Core™ Duo Processor

Efraim Rotem – Mobile Platform Group, Intel corporation
Jim Hermerding – Mobile platform Group, Intel corporation
Cohen Aviad - Microprocessor Technology Lab, Intel corporation
Cain Harel - Microprocessor Technology Lab, Intel corporation

**Abstract**

Modern CPUs with increasing core frequency and power are rapidly reaching a point where the CPU frequency and performance are limited by the amount of heat that can be extracted by the cooling technology.  In mobile environment, this issue is becoming more apparent, as form factors become thinner and lighter.  Often, mobile platforms trade CPU performance in order to reduce power and manage thermals. This enables the delivery of high performance computing together with improved ergonomics by lowering skin temperature and reducing fan acoustic noise.
Most of available high performance CPUs provide thermal sensor on the die to allow thermal management, typically in the form of analog thermal diode. Operating system algorithms and platform embedded controllers read the temperature and control the processor power. Improved thermal sensors directly translate into better system performance, reliability and ergonomics.
In this paper we will introduce the new Intel® Core™ Duo processor temperature sensing capability and present performance benefits measurements and results.

## Introduction

Today's high performance processors contain over a hundred million transistors, running in a frequency of several gigahertz. The power and thermal characteristics of these processors are becoming more challenging than ever before, and are likely to continue to grow with Moor's low. Improvements in the cooling technology however, are relatively slow and do not follow Moor's low. All computing segments face power and thermal challenges. In the server domain, the cost of electricity and air conditioning is one of the biggest expense items of a data center, and drives the need for low power high efficiency systems. In the mobile computing market, power and thermal management are the key limiter for delivering higher computational performance. Thin and light industrial designs are limited by the heat that can be extracted from the box. Ergonomic characteristics are also highly impacted by thermal considerations. The cooling fan is the major source of acoustic noise in the mobile system and external skin hot spots should be avoided for ergonomic reasons as well.

 The increasing demand for compute density brings the need for efficient thermal management schemes. Several such schemes have been proposed, for example DVS (dynamic frequency voltage scaling [1]). These mechanisms were implemented in CPUs such as the Intel® Centrino® Processor [2]. Most operating systems on the market support ACPI [3]. This is an industry standard infrastructure that enables thermal management of computer platforms. Thermal management is done by the use of active cooling devices, such as fans, or passive cooling actions such as DVS. Thermal management schemes accept user preferences for setting management policy. A computer user can select between high performance, energy conservation and improved ergonomics parameters.

The basic feedback for most of the power and thermal management schemes is temperature measurement. Both Intel® processors [4] and others [5], incorporate temperature sensor on the die to allow thermal measurement, typically in the form of analog thermal diode. The voltage on a diode junction is a function of the junction temperature. The diode is routed to external pins and an A/D chip on the platform converts the voltage into temperature reading. The Intel® Centrino processor [2] introduced a fixed thermal sensor, tuned to the max specified junction temperature. In case of abnormal conditions, such as cooling system malfunction, the circuit asserts a signal that activates a programmable self management power saving action that protects the CPU from operating out of its specified thermal range. It is apparent that the accuracy of the thermal measurements directly impacts the performance of the thermal management system and the performance of the CPU. In mobile computers, 1.5°C accuracy in temperature measurement is equivalent to 1 Watt of CPU power. In desk-top computers the impact is even higher due to the lower thermal resistance and 1°C accuracy translates into 2 Watt of CPU power.

There are several causes for temperature measurement inaccuracy:

1. Parameter variance: The thermal diode is not ideal and during the manufacturing process, there are variations in the diode parameters that translate into reading variations. An offset value is programmed into the Intel® Core™ Duo, to be used by the A/D to generate accurate readings.

2. A/D accuracy: Some errors are associated with the analog to digital conversion due to design and technology limitations as well as quantization errors. The best temperature A/Ds available on the market today provide +/- ½°C accuracy.

3. Proximity to the hot spot: CPU performance and reliability is limited by the temperature of the hottest location on the die. Thermal diode placement is limited by routing and I/O considerations and usually cannot be placed at the hottest spot on the die. Furthermore, the hot spot tends to shift around as a function of the workload of the CPU. It is not rare to find temperature difference as high as 10°C between a diode and the hot spot.

4. Manufacturing temperature control: Parts are tested for functionality and reliability at the max temperature specifications. Variations in test temperature drive a need for additional guard-band in the temperature control set points.

The speed of response to temperature changes also impacts thermal management performance. The Intel® Core™ Duo processor has implemented a new digital temperature reading capability to address the accuracy and response time limitations of existing solutions. The rest of the paper will describe the implementation of the digital sensor and the measured results of it's performance.

## The Intel® Core ™ Duo digital sensor (DTS)

The general structure of the digital thermometer of the Intel® Core™ Duo [7] is described in Figure 1. In addition to the analog thermal diode, multiple sensing devices are distributed on the die in all the hot spots. An internal A/D circuit converts each sensor into a 7 bit digital reading. The temperature reading is calculated as an offset from the maximum specified Tj, e.g. 0 indicated that the CPU is at it's maximum allowable Tj, 1 indicated 1°C below etc. All temperature readings are combined together into a single value, indicating the temperature of the hottest spot on die. The Intel® Core™ Duo is a dual core CPU. The DTS offers the ability to read temperature for each core independently and to read the maximum temperature for the entire package. To achieve measurement accuracy, each sensor is calibrated at test time. Calibration is done for the Maximum Tj and the linearity of the readout slope. The temperature reading is post processed for filtering out random noise and generating the H/W activated thermal protection functions. The DTS implementation on the Intel® Core™ Duo processor supports the legacy Intel Centrino® thermal sensor and fixed function thresholds PROCHOT and THERMTRIP [2].

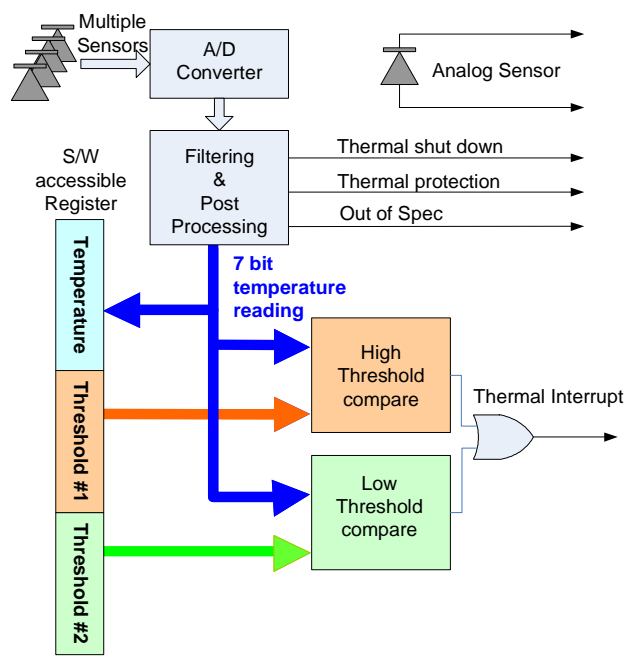

Figure 1: Digital Thermometer Block Diagram

PROCHOT is a fixed temperature threshold calibrated to trip at the max specified junction temperature. Upon crossing this threshold, a H/W power reduction action is initiated, reducing the frequency and voltage, keeping the CPU within functionality and reliability limits. A properly designed cooling system with thermal management should not activate the H/W protection mechanisms. Some aggressive platform designs however, may need occasional H/W initiated action due to long response time. On most operating systems, interrupt latency is not guaranteed and therefore, S/W based control may respond too slow. Other actions have inherent long delays. The time extending from activation of a fan and until its maximum speed is reached may be too long. Aggressive thermal design, together with a slow cooling response may cause thermal excursions that may compromise reliability and functionality. It is possible to design a system with enough margins to avoid such cases, but this comes at a cost of performance or compromised ergonomic characteristics. H/W based protection enables better user experience without compromising the device reliability and performance.

THERMTRIP is a catastrophic shut down event, both on the CPU and for the platform. It identifies thermal runaway in case of cooling system malfunction and turns off the CPU and platform voltages, preventing meltdown and permanent damage.

A new functionality of the DTS on the Intel® Core™ Duo is out of spec indication. It is possible for the CPU to operate within specifications while at maximum Tj. Out of spec indication is a notification to the operating system that a malfunction occurred, junction temperature is rising and a graceful shut down is required while functionality is still guaranteed and user data can be saved.

In order to perform S/W and ACPI thermal control functions, the DTS offers interrupt generation capability, in addition to the temperature reading. Two S/W programmable thresholds are loaded by S/W and a thermal interrupt is generated upon threshold crossing. This thermal event generates an interrupt to single or both cores simultaneously according to the APIC settings.

The digital thermometer is the basis for software thermal control such as the ACPI. In the ACPI infrastructure, thermal management is done by assigning a set of policies or actions to temperature thresholds. A policy can be active, such as activating fan in various speeds (_ACx), or passive (_PSV), by reducing the CPU frequency. Interrupt thresholds are defined to indicate upper and lower temperatures thresholds. An example of digital thermometer usage is given in Figure 2.

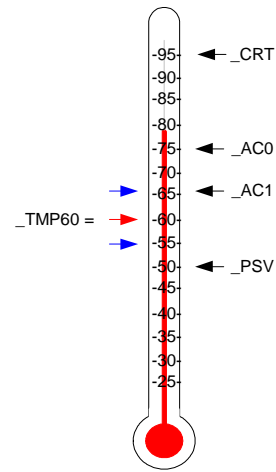

Figure 2: Digital thermometer and ACPI

In the above example the current die temperature is at $60^{o}C$. The thresholds set to $5^{o}C$ above and below the current temperature. If the temperature rises above $65^{o}C$, an interrupt is generated, notifying the S/W of a significant change in temperature. The control software reads the temperature and identifies the new temperature and initiates action if needed. In the above example, $65^{o}C$ requires activating a fan at a low speed. The activation thresholds and policies are defined at system configuration and communicated to the ACPI. Upon interrupt servicing, new thresholds are written around the new temperature to further track temperature changes. Small hysteresis values are applied to prevent frequent interrupts around a threshold point.

## Measurements and results

In previous Pentium™ - M systems, a single analog thermal diode was used to measure die temperature. Thermal diode cannot be located at the hottest spot of the die due to design limitations. To perform thermal management activities, some fixed offset was applied to the measured temperature, to keep the CPU within specifications. With the increasing performance and power density of the Intel® Core™ Duo, the performance implications of guard bands increase. Figure 3 shown measured die temperature of different workloads. It can be seen that the hot spot of the die moves to different locations depending on the nature of the workload.

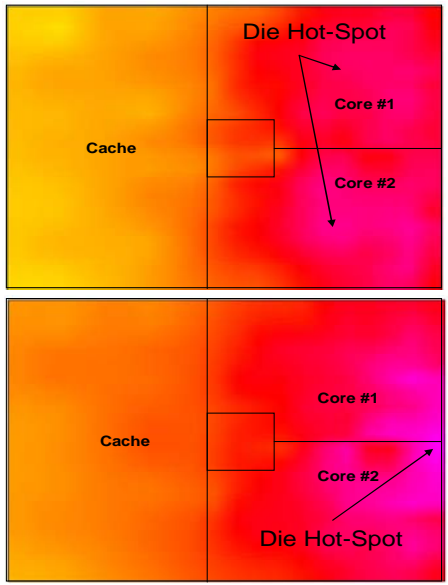

*Figure 3: Die hot spots at different workloads*

Figure 3 demonstrates a shift of the hot spot in a dual core workload. A workload that stresses the floating point unit which is a high power operation, will generate hot spot near the floating point while other workloads will stress different locations on the die. Figure 4 shows the thermal impact of single core applications.

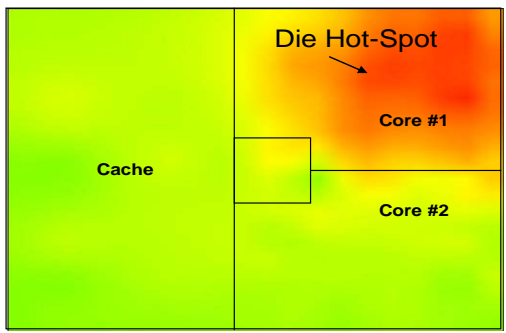

Figure 4: Thermal behavior of a single core application

It can be noted that a single diode cannot capture the maximal die temperature. Placing a diode between the cores, results in non optimal location as this is a relatively cold area of the die in single thread workloads. Workloads can be migrated by the operating system scheduler from one core to the another on the same die and therefore a symmetrical sensor placement is required.

In order to evaluate the DTS temperature reading, we performed a study to identify the impact of different workloads on the difference between diode and the hot spot, as measured by the DTS. A set of workloads including all SPEC-2K components and other popular benchmarks and applications, at single thread and multi-thread were executed on the CPU. Several iterations were done to reach a thermal steady state and then the diode and DTS temperatures were measured. Before taking the measurement, a calibration process has been performed, leaving only the temperature offset. As described earlier, both external A/D and internal DTS have some inaccuracies. Calibration procedure is needed to equalize DTS and diode temperature readings and measure temperature offsets only. Figure 5 shows the offset between the analog diode and the hot spot, as measured by the DTS. The horizontal axis represents the hot spot temperature as a percentage of the max temperature. The vertical axis shows the temperature offset between the diode and the hot spot. Each point on the chart represents a single application.

It can be seen that large temperature gradients exist on the die. It also can be noted that some workloads display high temperature gradients while other have no offset. Thermal control algorithms need to prevent the hot spot from exceeding the max temperature specification. It is possible to mitigate the temperature difference by applying a fixed offset to the diode reading. This obviously is a non optimal solution as the workloads with low offset will be panelized by the unnecessary temperature offset. The use of digital thermometer provides improved temperature reading, enables higher CPU performance within thermal limitations and improves reliability.

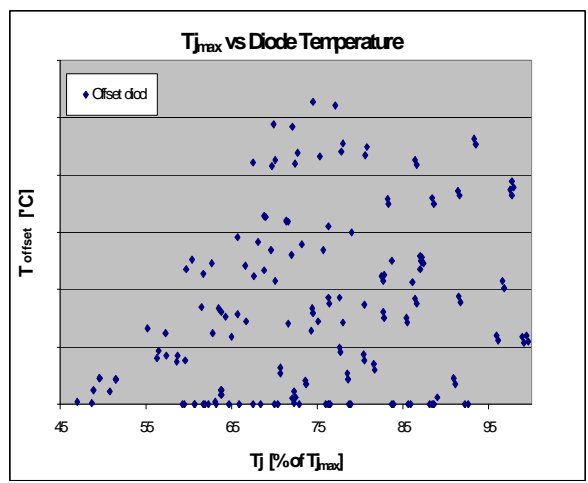

Figure 5: Diode to DTS Temp. difference

Previous studies [6] have shown that temperature reduction directly translates into performance degradation. The above chart represents 3%-7% reduction in performance due to temperature measurement offset. Building a thermal management system around a thermal diode, with the characteristics shown in Figure 5 requires temperature guard-band. This guard band can be applied to the control set point, and as a result, the workloads that generate high offset temperatures result in lost performance. A different approach can set the Tj threshold assuming that the diode represents the correct die temperature. Some of the workloads will run at high max Tj and therefore risk functional issues or reliability degradation.

The DTS also reduces the other temperature readings errors, which are not shown in this paper. The DTS is calibrated at manufacturing conditions and the reference point is set to this test temperature. Functionality, electrical specifications and reliability commitments are guaranteed at maximum Tj as measured by the DTS. Any test inaccuracy or parameters variance are already accounted for in the DTS set point.

## Summary and conclusions

With the increasing demand for computational density and the increase in CPU transistors and frequency, power and thermal are the key limiters for providing computing performance. In recent years, thermal management has become a fundamental function of computer platform. The input to every thermal management scheme is a thermal sensor. We have shown that thermal sensor accuracy translates into power, which in turn translates into better CPU performance. Legacy analog thermal sensors incur inaccuracies due to parameter distribution and temperature offset from the hot spot. In this paper we introduced the new digital thermal sensor (DTS) of the Intel® Core® Duo processor. We showed that multiple sense point on various hot spots of the die, together with on die A/D converter provide improved temperature reading. The better accuracy translates either into 3%-7% higher performance or into improved ergonomics. The introduction of dual core